\RequirePackage[2020-02-02]{latexrelease}
\documentclass[twocolumn,english,aps,prl,10pt,superscriptaddress]{revtex4}
\usepackage[T1]{fontenc}
\usepackage{amsmath,amssymb,graphicx}
\usepackage{float}
\usepackage{xcolor}
\begin{document}

\title{Mpemba Effect in Many-Body Systems Near Equilibrium}

\author{P. Ben-Abdallah}
\email{pba@institutoptique.fr}
\affiliation{Laboratoire Charles Fabry, UMR 8501, Institut d'Optique, CNRS, Universit\'e Paris-Saclay, 2 Avenue Augustin Fresnel, 91127 Palaiseau Cedex, France}

\date{\today}

\begin{abstract}
The Mpemba effect, in which a system initially farther from equilibrium
relaxes faster than a closer one, has been observed in a wide variety of
linear and nonlinear systems. Here we develop a unified framework for the
Mpemba effect in many-body systems near equilibrium based on the spectral
geometry of the relaxation operator. We distinguish a non-uniform Mpemba effect,
associated with a crossing of global distances to equilibrium, from a strict
componentwise Mpemba effect, in which the initially hotter state remains
larger in every degree of freedom yet relaxes faster. We show that reciprocal
systems admit only the former, whereas reciprocity breaking renders the
relaxation operator non-normal and can enable the latter. These results
identify reciprocity and non-normality as key ingredients governing anomalous
relaxation in linear many-body systems.
\end{abstract}

\maketitle
\section{Introduction}

\begin{figure}
\centering
\includegraphics[angle=0,scale=0.35,angle=0]{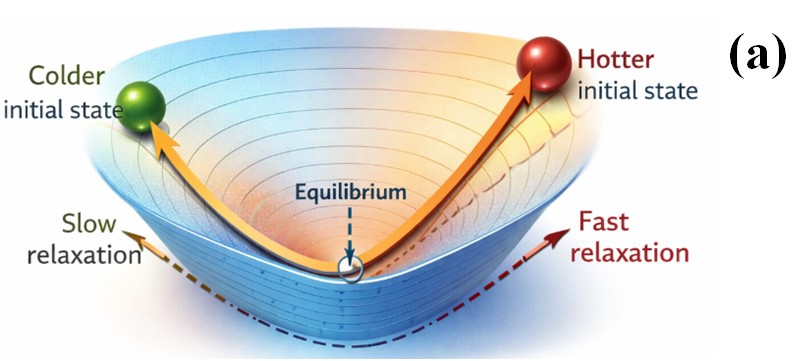}
\includegraphics[angle=0,scale=0.38,angle=0]{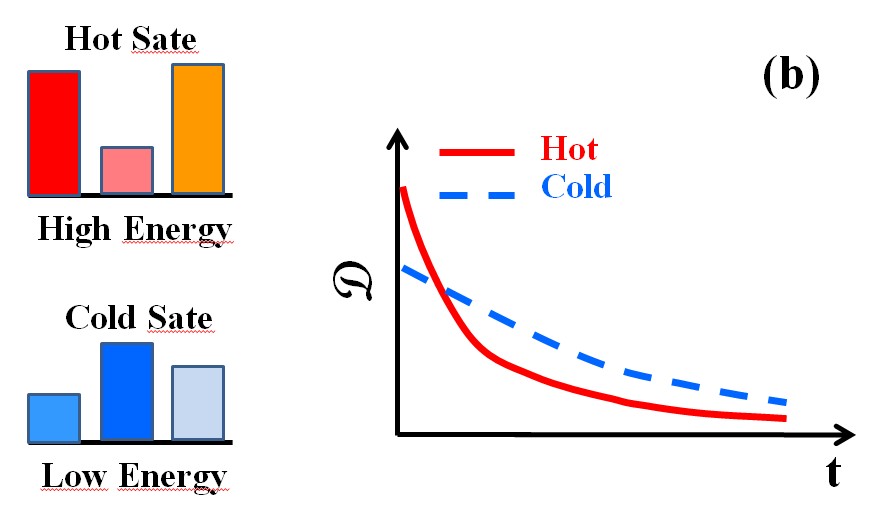}
\includegraphics[angle=0,scale=0.38,angle=0]{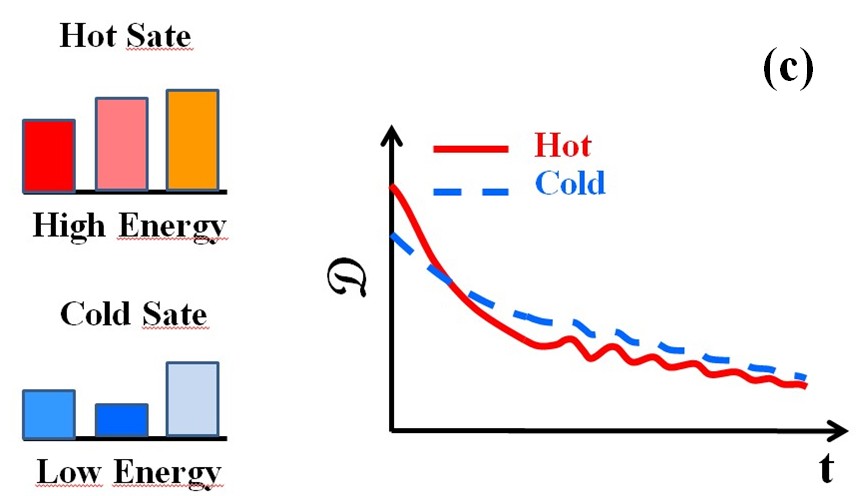}
\caption{(a) Mechanical analogy of the Mpemba effect: relaxation corresponds to overdamped motion in an effective potential $V_{\mathrm{eff}}(\Theta)=\tfrac12 \Theta^T M \Theta$, with the bottom representing equilibrium. Two initial states are shown: a hotter state (red) aligned with steep directions, and a colder state (green) along shallow directions. (b) Non-uniform Mpemba effect with a many-body system: the hotter state relaxes faster globally without being componentwise larger. (c) True Mpemba effect with a many-body system: the hotter state is strictly larger in all components and still relaxes faster.}
 \label{fig Mpemba}
\end{figure}

When two systems relax toward a common equilibrium, one naturally expects that the state initially farther from equilibrium will remain farther away at all subsequent times. Relaxation processes typically preserve the ordering of deviations from equilibrium. Surprisingly, under certain conditions a system that starts farther from equilibrium can approach equilibrium faster than one that begins closer. This counterintuitive phenomenon is known as the \textit{Mpemba effect}, first reported in water freezing experiments~\cite{Mpemba} and subsequently observed in a wide range of classical and quantum systems~\cite{Lasanta,Kumar,Lu,Klich,Ares,Baity}. In general, a Mpemba effect occurs when the global distance to equilibrium of an initially hotter system crosses that of a colder one at a finite time (Fig.~\ref{fig Mpemba}).
The Mpemba effect has been investigated in systems ranging from granular fluids and spin glasses to stochastic Markov processes, radiative thermal networks, and quantum systems~\cite{Lasanta,Lu,Klich,Baity,Ares}. Existing theoretical descriptions reveal several mechanisms underlying anomalous relaxation, including metastability, nonlinear energy landscapes, and spectral properties of linear relaxation operators. In particular, linear Markovian theories have shown that a hotter state may relax faster when its projection onto the slowest relaxation mode is sufficiently suppressed~\cite{Lu,Klich}. A similar mechanism has recently been identified in many-body radiative heat-transfer systems operating within the linear-response regime. Despite these advances, a unified understanding of how reciprocity and many-body interactions constrain the different manifestations of the Mpemba effect remains lacking.
In this work, we revisit the Mpemba effect within the linear-response regime of many-body systems and establish a unified geometric framework based on the spectral properties of the relaxation operator. Rather than focusing on the existence of anomalous relaxation in linear dynamics, which is already known in both Markovian~\cite{Lu,Klich} and radiative systems~\cite{Messina2013,PRL2021}, we identify the conditions under which different forms of the Mpemba effect can occur. Specifically, we distinguish a non-uniform Mpemba effect, characterized by a crossing of global distances to equilibrium without componentwise ordering, from a strict componentwise (uniform) Mpemba effect, where the initially hotter state remains larger in every degree of freedom yet relaxes faster globally. We show that reciprocity imposes strong constraints on anomalous relaxation: reciprocal systems admit only the former, whereas nonreciprocal systems may exhibit the latter through the non-normal geometry of their relaxation operators. Our results therefore provide a classification of Mpemba phenomena in terms of the spectral geometry of linear many-body dynamics.

\section{General framework for linear many-body relaxation}

We consider a system with $N$ degrees of freedom relaxing toward a common equilibrium state. The vector
\begin{equation}
\Theta=(\Theta_1,\ldots,\Theta_N)^{\top}\in\mathbb{R}^N
\end{equation}
denotes the deviations of these degrees of freedom from equilibrium. In the linear-response regime, the dynamics are governed by
\begin{equation}
\dot{\Theta}=-M\Theta ,
\label{dynamics}
\end{equation}
where $M$ is the relaxation matrix.

\subsection{Uniform and non-uniform Mpemba effects}

Throughout this work, we distinguish between two forms of the Mpemba effect, referred to as uniform and non-uniform. In both cases, the initially hotter state is assumed to be farther from equilibrium than the colder one according to the chosen global distance measure,
\begin{equation}
\|\Theta^{(h)}(0)\|>\|\Theta^{(c)}(0)\|.
\end{equation}
A uniform Mpemba effect occurs when the hotter state also dominates the colder state componentwise,
\begin{equation}
\Theta_i^{(h)}(0)>\Theta_i^{(c)}(0),
\qquad \forall i,
\end{equation}
yet nevertheless relaxes faster and becomes closer to equilibrium at a finite time.
By contrast, a non-uniform Mpemba effect occurs when this componentwise ordering is not satisfied, although the hotter state still has the larger initial distance from equilibrium. We emphasize that the terminology ``uniform'' and ``non-uniform'' refers exclusively to the componentwise ordering of the initial states and not to their global distance from equilibrium, which is larger for the hotter state in both cases.

\medskip
\section{Reciprocal many-body systems}

\subsection{Modal decomposition and relaxation geometry}
 
When microscopic reciprocity holds, the relaxation matrix $M$ is
symmetric and positive definite, with non-positive off-diagonal entries.
The dynamics defined by Eq.~(\ref{dynamics}) can then be decomposed into independent
collective modes,
\begin{equation}
\Theta(t) = \sum_{k=1}^{N} a_k v_k e^{-\lambda_k t},
\end{equation}
with eigenvectors $v_k$ and rates $\lambda_k$. Large $\lambda_k$ correspond to fast-decaying modes, small $\lambda_k$ to slow bottlenecks. The initial projections $a_k$ determine the global relaxation behavior.

\subsection{Absence of Mpemba effects in reciprocal two-body systems}

For two-body systems, all eigenvector components have the same sign. Consequently, the initial ordering of modal amplitudes is preserved,
\begin{equation}
a_\pm^{(h)} - a_\pm^{(c)} \ge 0,
\end{equation}
and the global norm difference
\begin{align}
\Phi(t) &= \|\Theta^{(h)}(t)\|^2 - \|\Theta^{(c)}(t)\|^2 \notag\\
        &=A_+ e^{-2 \lambda_+ t} + A_- e^{-2 \lambda_- t}, \quad A_\pm \ge 0,
\end{align}
cannot change sign. Therefore, neither a non-uniform nor a uniform Mpemba effect can occur in reciprocal two-body systems.

\subsection{Non-uniform Mpemba effects in reciprocal many-body systems}

For three or more degrees of freedom, the situation changes because the eigenvectors of the relaxation matrix can have mixed signs. As a result, the hotter state may project more strongly onto fast-decaying modes and more weakly onto slow ones. In that case, the difference in global distance to equilibrium
\begin{equation}
\Phi(t) = \sum_k \Delta a_k \, e^{-2 \lambda_k t}
\end{equation}
can change sign at a finite time $t^\ast$, producing a Mpemba effect.
For reciprocal systems, the matrix $-M$ is a Metzler matrix, since
$M_{ii}>0$ and $M_{ij}\le 0$ for $i\neq j$.
Consequently, the semigroup
$e^{-Mt}$
is positivity preserving
\begin{equation}
\Theta(0)\ge 0
\quad\Longrightarrow\quad
\Theta(t)\ge 0,
\qquad \forall t>0.
\label{eq:positivity}
\end{equation}
More generally, if two initial conditions satisfy
\begin{equation}
\Theta^{(h)}(0)-\Theta^{(c)}(0)\ge 0
\end{equation}
componentwise, then
\begin{equation}
\Theta^{(h)}(t)-\Theta^{(c)}(t)
=
e^{-Mt}
\bigl[\Theta^{(h)}(0)-\Theta^{(c)}(0)\bigr]
\ge 0.
\label{eq:ordering}
\end{equation}
for all times.
Therefore, the positivity-preserving property of $e^{-Mt}$ guarantees that componentwise ordering is maintained throughout the relaxation process. As a consequence, a strict componentwise (uniform) Mpemba effect is impossible in reciprocal linear systems. The only admissible form of anomalous relaxation is a non-uniform Mpemba effect, in which the initially hotter state relaxes faster in terms of its global distance to equilibrium despite not dominating the colder state componentwise. In this case, the crossing originates from a redistribution of the initial condition among fast and slow relaxation modes, allowing the hotter state to carry a comparatively smaller weight in the slowest mode.

Geometrically, the relaxation can be viewed as a gradient descent in the quadratic potential
\begin{equation}
V_{\rm eff}(\Theta) = \tfrac12 \Theta^T M \Theta, 
\qquad 
\dot{\Theta} = - \nabla_\Theta V_{\rm eff}.
\end{equation}
In this picture (Fig.\ref{fig Mpemba}), fast modes correspond to steep directions of the potential landscape, while slow modes correspond to shallow directions. A hotter initial state can therefore relax faster globally if it is preferentially aligned with the steep directions, even though the componentwise ordering of the states is preserved.

The mechanism described above is closely related to the Mpemba index
introduced by Klich \textit{et al.}~\cite{Klich} in the context of
Markovian relaxation. In that framework, a Mpemba effect occurs when the
initially hotter state possesses a smaller projection onto the slowest
relaxation mode than the colder state. The non-uniform Mpemba effect discussed
here recovers the same spectral mechanism in the language of linear
many-body relaxation. Our formulation generalizes this perspective beyond
Markov generators and highlights the role of reciprocity and non-normality.
In particular, while the reciprocal case reduces to the familiar
slow-mode suppression mechanism, nonreciprocal systems exhibit additional
geometric effects associated with the misalignment of left and right
eigenvectors, which can lead to a strict componentwise (uniform) Mpemba effect.

\begin{figure}[t]
    \centering
    \includegraphics[width=0.48\textwidth]{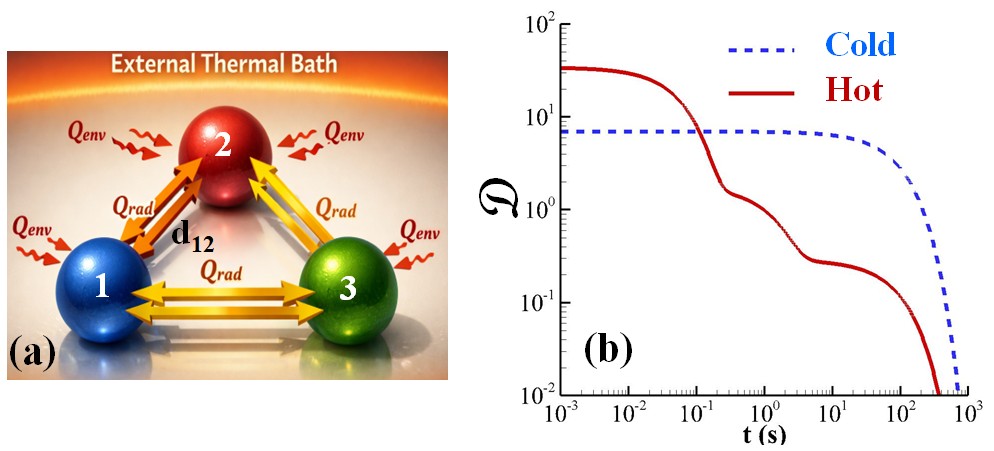} 
    \caption{%
    Mpemba effect in a many-body system (a) made with three SiC nanoparticles (radius $r = 50~\mathrm{nm}$) arranged in a scalene triangle which exhange heat between them and with an external bath by radiation in near-field and far-field, respectively. Double arrows indicate the strength of radiative coupling $G_{ij}$, the shortest separation $d_{12}=3r$ corresponding to the strongest interaction. All particles are coupled to the bath (wavy arrows) at $T_b = 300~\mathrm{K}$. (b) Distance to the equilibrium $\mathcal{D}(t)=\|\Theta^{(h,c)}(t)\|$ for a ``hot'' initial state $\Theta^{(h)}(0)=(25,-23,-1.5)$ (red) and a ``cold'' initial state $\Theta^{(c)}(0)=(4,4,4)$ (blue). Despite starting farther from equilibrium, the hot state relaxes faster due to suppressed projection onto the slowest mode, producing a clear crossing, showing a non-uniform Mpemba effect.
    }
    \label{fig:mpemba_three_body}
\end{figure}
\medskip

\section{Nonreciprocal systems}

\subsection{Non-normal relaxation operators}

The nonreciprocal relaxation operators considered here are closely related,
at a mathematical level, to non-detailed-balance Markov generators
investigated in previous studies of the Mpemba effect~\cite{Lu}. In both
cases the relaxation operator is generally non-symmetric and the dynamics
cannot be described by a gradient flow. The present framework extends this
perspective beyond stochastic dynamics to general linear many-body systems,
including thermal, mechanical, electrical and active networks. This broader
setting naturally highlights the role of non-normality and the associated
biorthogonal geometry of left and right eigenvectors in shaping anomalous
relaxation.
When reciprocity is broken,  the relaxation matrix $M$ is no longer symmetric. Although stability still requires $\mathrm{Re}\,\lambda_k>0$ for all eigenvalues, the operator is generally non-normal, $MM^\dagger \neq M^\dagger M$. Then, right and left eigenvectors differ and form a biorthogonal basis.
The propagator admits the spectral decomposition
\begin{equation}
e^{-M t} = \sum_k e^{-\lambda_k t}\, v_k \otimes w_k,
\qquad
w_k^{\!\top} v_\ell = \delta_{k\ell},
\end{equation}
so that modal amplitudes are determined by projections onto the left eigenvectors,
\begin{equation}
a_k = w_k^{\!\top}\Theta(0),
\end{equation}
rather than by orthogonal projections as in reciprocal systems. Because left and right eigenvectors are misaligned, non-normal dynamics can transiently rotate and shear state space even though all eigenvalues correspond to decay.
As a consequence, ordering of initial Euclidean norms does not necessarily imply the ordering of modal amplitudes. An initially hotter configuration may carry a larger weight in fast-decaying directions while being comparatively depleted along slow subspaces, producing accelerated short-time relaxation. The difference vector $\Delta(t)=e^{-Mt}\Delta(0)$ need not remain aligned with $\Delta(0)$, reflecting directed probability or energy currents induced by asymmetric couplings.

However, for cooperative systems the situation is more constrained. The
condition $M_{ij}<0$ ($i\neq j$) is the natural sign structure of passive
diffusive couplings. Indeed, thermal, electrical, and overdamped mechanical
networks can generally be written as
\begin{equation}
\dot\Theta_i
=
-G_i\Theta_i
+\sum_{j\neq i}K_{ij}(\Theta_j-\Theta_i),
\qquad K_{ij}>0,
\end{equation}
which is equivalent to $\dot\Theta=-M\Theta$ with
\begin{equation}
M_{ii}=G_i+\sum_{j\neq i}K_{ij}>0,
\qquad
M_{ij}=-K_{ij}<0 \quad (i\neq j).
\end{equation}
A positive perturbation of component $j$ increases the growth rate of
component $i$, so that the off-diagonal influence is positive in the
sense of monotone dynamical systems. This sign structure defines a
cooperative system.

\subsection{Cooperative two-body systems}

For a two-dimensional cooperative nonreciprocal system,
\begin{equation}
M=
\begin{pmatrix}
a & -b\\
-c & d
\end{pmatrix},
\qquad
a,d>0,\quad b,c>0.
\label{eq:2Dmatrix}
\end{equation}
the slow left eigenvector $w_s=(w_1,w_2)^T$ satisfies
\[
(a-\lambda_s)w_1-cw_2=0,
\qquad
-bw_1+(d-\lambda_s)w_2=0,
\]
where $\lambda_s$ is the smallest eigenvalue. Since
\begin{equation}
\lambda_s=
\frac{a+d-\sqrt{(a-d)^2+4bc}}{2}
<\min(a,d).
\label{eq:sloweigenvalue}
\end{equation}
it follows that
\begin{equation}
\frac{w_2}{w_1}
=
\frac{a-\lambda_s}{c}
>0,
\qquad
\frac{w_1}{w_2}
=
\frac{d-\lambda_s}{b}
>0.
\label{eq:leftpositive}
\end{equation}
Hence the slow left eigenvector can be chosen strictly positive.
If two initial states satisfy
\[
\Theta^{(h)}(0)-\Theta^{(c)}(0)>0
\]
componentwise, then their slow-mode amplitudes obey
\begin{equation}
a_s^{(h)}-a_s^{(c)}
=
w_s^{T}
\left[
\Theta^{(h)}(0)-\Theta^{(c)}(0)
\right]
>0.
\label{eq:slowmodeordering}
\end{equation}
The hotter state therefore necessarily carries a larger weight in the
slowest relaxation mode. Since the asymptotic dynamics are controlled by
this mode, the ordering cannot be inverted at long times. Consequently,
although non-normality can already produce a non-uniform Mpemba effect in
two dimensions, a strict componentwise (uniform) Mpemba effect remains excluded for
cooperative two-body systems. By contrast, in higher dimensions, or in active and non-cooperative systems
where left eigenvectors may contain components of different signs, the
ordering of slow-mode amplitudes is no longer constrained by componentwise
ordering of the initial conditions. Reciprocity breaking can then invert the
slow-mode hierarchy and produce a genuine componentwise Mpemba effect within
linear dynamics.

\subsection{Conditions for a uniform Mpemba effect}

With the modal decomposition
\begin{equation}
\Theta(t)=\sum_k a_k v_k e^{-\lambda_k t},
\end{equation}
the squared distance to equilibrium is
\begin{equation}
\|\Theta(t)\|^2
=\sum_{k,l} a_k a_l \,(v_k\cdot v_l)\,
e^{-(\lambda_k+\lambda_l)t}.
\end{equation}
For two initial states, $h$ (hot) and $c$ (cold), the global norm difference reads
\begin{equation}
\Phi(t)=
\sum_{k,l}
\left(a_k^{(h)}a_l^{(h)}-a_k^{(c)}a_l^{(c)}\right)
(v_k\cdot v_l)
e^{-(\lambda_k+\lambda_l)t}.
\end{equation}
Thus a true Mpemba effect occurs if and only if $\Phi(t)$ changes sign at a finite time $t^\ast$, which requires
\begin{equation}
\sum_{k,l}
\left(a_k^{(h)}a_l^{(h)}-a_k^{(c)}a_l^{(c)}\right)
(v_k\cdot v_l)
e^{-(\lambda_k+\lambda_l)t^\ast}=0 .
\end{equation}
The above expression for $\Phi(t)$ shows that it is a finite sum of exponentials with
different decay rates. A necessary condition for a Mpemba crossing is that
the coefficients
\begin{equation}
C_{kl}=
\left(a_k^{(h)}a_l^{(h)}-a_k^{(c)}a_l^{(c)}\right)
(v_k\!\cdot\! v_l).
\label{eq:Ckl}
\end{equation}
are not all of the same sign. If all $C_{kl}$ are positive (or all negative),
$\Phi(t)$ remains sign-definite and no crossing can occur.
Conversely, when positive and negative contributions coexist, fast-decaying
terms can dominate at short times whereas slow terms control the asymptotic
dynamics. In this case $\Phi(t)$ may change sign at a finite time,
producing a Mpemba effect. The crossing therefore originates from the
competition between modal contributions associated with distinct relaxation
rates.A sufficient condition follows from continuity.
If $\Phi(0)>0$ and the coefficient associated with the
slowest-decaying contribution is negative, then
$\lim_{t\to\infty}\Phi(t)<0$ and at least one finite-time
crossing must exist.

\section{Physical examples}

\subsection{Radiative heat transfer in a three-body thermal network}

To finish, let us illustrate these results with two examples. 

\begin{figure}[t]
    \centering
    \includegraphics[width=0.48\textwidth]{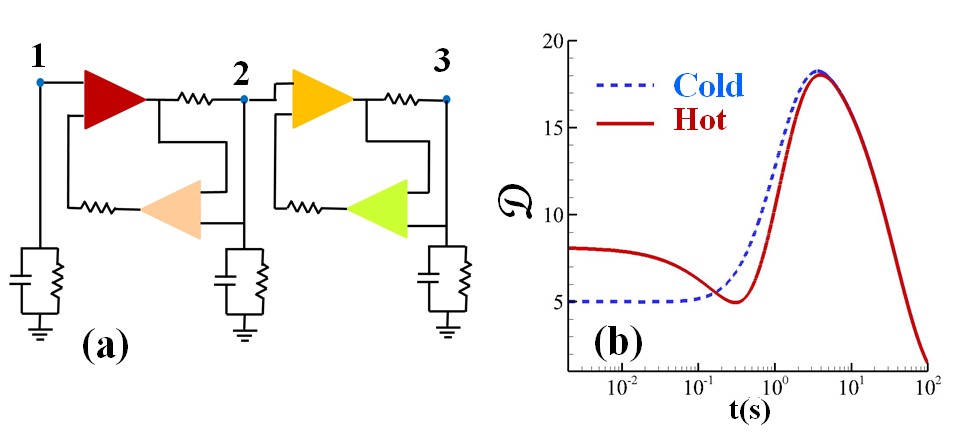} 
\caption{True Mpemba effect in an active many-body system.
(a) The system is an electronic circuit composed of three aligned nodes whose voltages represent the dynamical variables. Each node is connected to ground by a capacitor and a resistor in parallel, setting the local relaxation rates $M_{11}=0.1$, $M_{22}=1.0$, and $M_{33}=4.0$. Nodes $1$ and $2$ are coupled by two operational amplifiers arranged in parallel and in opposite directions; each op-amp injects a current through a series resistor before reaching the receiving node, thereby implementing directional couplings. The same architecture connects nodes $2$ and $3$. This design realizes asymmetric interactions with strengths $M_{12}=3.5$, $M_{21}=0.02$, $M_{23}=2.5$, $M_{32}=0.03$, while $M_{13}=M_{31}=0$.
(b) Time evolution of the distance $\mathcal{D}$ for the three coupled nodes, starting from two initial states in which the hot state is strictly larger componentwise than the cold state. The asymmetry of the couplings aligns the hot state with faster decaying modes, leading to a finite-time crossing of the Euclidean norms and providing evidence of a genuine componentwise Mpemba effect.}
    \label{fig:true mpemba}
\end{figure}

\medskip

Consider three identical silicon carbide nanoparticles~\cite{Palik} of radius $r = 50\,\mathrm{nm}$ and heat capacity $C = 1.1\times10^{-15}\,\mathrm{J/K}$, modeled as point dipoles with polarizability $\alpha(\omega)$, embedded in a thermal bath at $T_b = 300\,\mathrm{K}$. Let $T_i(t)$ denote the temperature of particle $i$ and define deviations $\Theta_i = T_i - T_b$, with $\Theta = (\Theta_1,\Theta_2,\Theta_3)^T$.
Within fluctuational electrodynamics~\cite{Biehs2021}, the radiative power exchanged between particles $i$ and $j$ (including the bath) is
\begin{equation}
\mathcal{P}_{i\leftrightarrow j}
= \int_0^\infty \frac{d\omega}{2\pi}
\,\hbar \omega\, \tau_{ij}(\omega)\,
\Delta n_{ij}(\omega,T),
\end{equation}
where $\tau_{ij}(\omega)=4 \omega^4/c^4\,\mathrm{Im}[\alpha]^2\,\mathrm{Tr}[\mathbb{G}_{ij}\mathbb{G}_{ij}^\dagger]$~\cite{pba2011} and $\Delta n_{ij} = n(\omega,T_i)-n(\omega,T_j)$ with $n(\omega,T)$ the Bose-Einstein distribution. The tensor $\mathbb{G}_{ij}$ is the full Green tensor including multiple scattering processes. Linearizing around $T_b$ defines conductances $G_{ij} = \partial \mathcal{P}_{i\leftrightarrow j}/\partial T_j|_{T_b}$. In the near-field dipolar regime, $G_{ij} \propto d_{ij}^{-6}$. The radiative conductance to the bath is estimated in the grey-body dipole approximation,
\begin{equation}
G_{b} \approx 4 \pi r^2 \, \varepsilon_{\mathrm{eff}} \, 4 \sigma T_b^3,
\end{equation}
with $\sigma$ the Stefan--Boltzmann constant and $\varepsilon_{\mathrm{eff}}$ the particle emissivity (with $\varepsilon_{\mathrm{eff}}\approx 0.8$, $G_b \sim 10^{-17}\,\mathrm{W/K}$). 
Then, the components of the dynamic matrix read
\begin{equation}
M_{ij}=\frac{1}{C}\left[\left(G_b+\sum_{k}G_{ik}\right)\delta_{ij}-G_{ij}\right].
\end{equation}
 To create a hierarchy of relaxation channels, particles 1 and 2 are placed at the minimal separation compatible with the dipolar approximation $d_{12} = 3r = 150\,\mathrm{nm}$.
yielding $G_{12} \sim 8\times 10^{-15}\,\mathrm{W/K}$, while $G_{13}=3\times 10^{-16}\,\mathrm{W/K}$, $G_{23}=1.5\times 10^{-16}\,\mathrm{W/K}$ define a weaker coupling for particle 3. Using the near-field scaling $G_{ij} \propto d_{ij}^{-6}$,
the corresponding interparticle distances $d_{ij} = d_{12} \left(\frac{G_{12}}{G_{ij}}\right)^{1/6}$ are determined. These unequal separations define a scalene triangular configuration, essential to avoid degenerate relaxation modes that would suppress the Mpemba effect. 

In Fig.\ref{fig:mpemba_three_body} we show the relaxation dynamics of this system for the two initial conditions
\begin{equation}
\Theta^{(h)}(0)=(25,-23,-1.5),\qquad
\Theta^{(c)}(0)=(4,4,4).
\end{equation}
Although the hot state starts with a significantly larger initial distance to equilibrium, its projection onto the slowest relaxation mode is strongly suppressed. As a result, it is dominated by fast-decaying modes and relaxes rapidly at short times.
In contrast, the cold initial condition has a smaller initial norm but a strong projection onto the slowest eigenmode. Consequently, its decay is governed by the slowest relaxation rate and remains elevated for longer times. This modal imbalance leads to a clear crossing of the two relaxation curves: the initially hotter system cools faster and becomes colder than the initially colder one, demonstrating a pronounced Mpemba effect.
The chosen initial conditions are not unique. The Mpemba effect persists
under small perturbations of the initial state, provided the projection
onto the slowest relaxation mode remains sufficiently suppressed compared
with that of the colder configuration. The reported state should therefore
be regarded as a representative point within a finite region of the
initial-condition space exhibiting the same behavior.

\medskip

\subsection{Nonreciprocal oscillator network}

A minimal realization of a three-dimensional Mpemba effect can be obtained in a linear system of three coupled overdamped modes with state vector $V=(V_1,V_2,V_3)^\top$. The dynamics obey
\begin{equation}
\dot V = - M V, \qquad
M_{ij} = G_i \, \delta_{ij} - G_{ij} \, (1-\delta_{ij}),
\end{equation}
where $G_i>0$ represents intrinsic damping (mechanically $G_i=k_i/\gamma_i$, electrically $G_i=1/R_i$). The off-diagonal terms $G_{ij}$ describe couplings between components: asymmetry ($G_{ij}\neq G_{ji}$) introduces nonreciprocity, while negative couplings represent non-cooperative or inhibitory interactions. In active realizations, effective negative $G_i$ may also arise, corresponding to elements that inject energy into the system.
Asymmetric couplings render the dynamical matrix generally non-normal,
while dissipation ensures the stability of the relaxation dynamics. As a result, the directions along which the system relaxes differ from those that determine how an initial state projects onto the dynamical modes. This geometric mismatch enables a genuine componentwise Mpemba effect: even when the “hot” initial state is strictly larger than the “cold” state in every component, the hot state can relax faster because its projection is biased toward rapidly decaying modes.

Figure~\ref{fig:true mpemba} illustrates this phenomenon in a minimal electrical implementation of three coupled overdamped modes. In the circuit, the dynamical variables are the node voltages $V_i(t)$. Each node is connected to ground through a resistor and a capacitor in parallel, producing passive relaxation, while directed couplings between nodes are implemented using operational amplifiers that inject currents through series resistors. This architecture generates a non-symmetric dynamical matrix $M$ with positive diagonal elements and asymmetric off-diagonal couplings, combining dissipative relaxation with nonreciprocal and weakly inhibitory interactions. The “hot’’ initial voltage configuration,
$V_h(0)=(5.1,\,6.0,\,2.0)^{\top}$ is strictly larger componentwise than the “cold’’ configuration $V_c(0)=(5.0,\,0.2,\,0.1)^{\top}$.
Despite this ordering, the hot state relaxes faster at long times. The asymmetry of the couplings aligns the hot initial condition predominantly with fast-decaying collective modes of the circuit, whereas the cold state has a larger projection onto slower modes associated with the smallest relaxation rates. 
Physically, nonreciprocity skews the relaxation geometry: decay directions (right eigenvectors) differ from projection directions (left eigenvectors). In combination with non-cooperative interactions or activity, this structure allows rapid decay along selected modes even under strict componentwise dominance. The resulting biorthogonal dynamics thus provides a minimal linear mechanism for a genuine three-dimensional Mpemba effect, in which all components of the initially hotter state are larger yet the system relaxes faster than the colder one.
Likewise, the strict componentwise Mpemba effect is not restricted to a
single pair of initial conditions. It persists for finite perturbations of
the hot and cold states as long as the ordering of the relevant modal
amplitudes is preserved. The example shown in Fig.~3 is therefore intended
to illustrate the mechanism rather than a fine-tuned isolated case.

\section{Discussion}

The present work places several previously identified manifestations of the Mpemba effect within a common spectral framework. In reciprocal systems, anomalous relaxation is governed by the distribution of the initial state among the relaxation eigenmodes and is closely related to the suppression of the slowest mode embodied in the Mpemba index introduced by Klich \textit{et al.}~\cite{Klich}. From this perspective, the non-uniform Mpemba effect emerges naturally when the initially hotter state carries a comparatively smaller weight in the slowest relaxation channel, despite being farther from equilibrium overall.

Breaking reciprocity qualitatively modifies this picture. In nonreciprocal systems, the relaxation operator becomes non-normal and its left and right eigenvectors no longer coincide. Consequently, the directions that determine how initial conditions are decomposed into modes differ from the directions along which perturbations decay. This biorthogonal geometry introduces a mechanism for anomalous relaxation that is absent in reciprocal systems and provides a natural interpretation of Mpemba effects in non-detailed-balance dynamics. While reciprocal systems can exhibit only non-uniform Mpemba behavior, nonreciprocal systems are not constrained by positivity-preserving dynamics and may therefore display a genuine uniform Mpemba effect, in which the hotter state remains larger in every degree of freedom yet relaxes faster globally.

More generally, our results identify reciprocity, non-normality and the sign structure of slow relaxation modes as the key ingredients governing Mpemba phenomena in linear many-body systems. The examples presented here are intended to illustrate these general principles rather than isolated realizations. The occurrence of both uniform and non-uniform Mpemba effects is robust to finite perturbations of the initial conditions, provided that the hierarchy of modal amplitudes responsible for the crossing is preserved. These findings suggest that the classification developed here applies broadly to thermal networks, electrical circuits, mechanical oscillator arrays, active matter systems, and stochastic relaxation processes.

\section{Conclusions}

To conclude, we have shown that the Mpemba effect can emerge entirely within the linear-response regime of many-body systems and is fundamentally governed by the spectral geometry of the relaxation operator. In reciprocal systems, relaxation follows a gradient flow: non-uniform Mpemba effects may occur in systems with three or more degrees of freedom through an appropriate redistribution of modal amplitudes, whereas a uniform (strict componentwise) Mpemba effect is excluded by the positivity-preserving structure of the dynamics. By contrast, breaking reciprocity renders the relaxation operator non-normal, so that projection and decay directions no longer coincide. This biorthogonal geometry provides a natural mechanism for anomalous relaxation beyond the constraints of reciprocal dynamics
and can produce a genuine uniform Mpemba effect.
More broadly, our results establish reciprocity, non-normality and the sign structure of relaxation modes as the key ingredients controlling Mpemba phenomena in linear many-body systems. They provide a unified framework encompassing reciprocal, nonreciprocal, passive, and active systems, and reveal how the interplay between spectral properties and relaxation geometry governs anomalous thermalization and equilibration processes.

\begin{acknowledgements}
Data availability. The data are available from the authors upon reasonable request.
\end{acknowledgements}


\begin{thebibliography}{99}
\bibitem{Mpemba}
E. B. Mpemba and D. G. Osborne,
Cool?,
Phys. Educ. \textbf{4}, 172 (1969).

\bibitem{Lasanta}
A. Lasanta, F. Vega Reyes, A. Prados, and A. Santos,
When the hotter cools more quickly: Mpemba effect in granular fluids,
Phys. Rev. Lett. \textbf{119}, 148001 (2017).

\bibitem{Kumar}
A. Kumar and J. Bechhoefer,
Exponentially faster cooling in a colloidal system,
Nature \textbf{584}, 64 (2020).

\bibitem{Lu}
Z. Lu and O. Raz,
Nonequilibrium thermodynamics of the Markovian Mpemba effect,
Proc. Natl. Acad. Sci. U.S.A. \textbf{114}, 5083 (2017).

\bibitem{Klich}
I. Klich, O. Raz, O. Hirschberg, and M. Vucelja,
Mpemba index and anomalous relaxation in Markovian dynamics,
Phys. Rev. X \textbf{9}, 021060 (2019).


\bibitem{Ares}
F. Ares, P. Calabrese and S. Murciano,
\textit{The quantum Mpemba effects},
Nature Reviews Physics \textbf{7}, 451–460 (2025)


\bibitem{Baity}
M. Baity-Jesi \textit{et al.},
The Mpemba effect in spin glasses,
Proc. Natl. Acad. Sci. U.S.A. \textbf{116}, 15350 (2019).

\bibitem{Messina2013}
R.Messina, M. Tschikin, S.-A. Biehs and P.  Ben-Abdallah, Fluctuation-electrodynamic theory and dynamics of heat transfer in systems of multiple dipoles, 
Phys. Rev. B \textbf{88}, 104307 (2013).

\bibitem{PRL2021}
S. Sanders, L. Zundel, W.J.M. Kort-Kamp, D.A.R. Dalvit and A. Manjavacas, Near-Field Radiative Heat Transfer Eigenmodes,
Phys. Rev. Lett. \textbf{126}, 193601 (2021).

\bibitem{VanKampen}
N. G. van Kampen,
\textit{Stochastic Processes in Physics and Chemistry}
(Elsevier, Amsterdam, 2007).

\bibitem{Trefethen}
L. N. Trefethen and M. Embree,
\textit{Spectra and Pseudospectra: The Behavior of Non-Normal Matrices and Operators}
(Princeton University Press, Princeton, 2005).

\bibitem{Schmid}
P. J. Schmid,
Nonmodal stability theory,
Annu. Rev. Fluid Mech. \textbf{39}, 129 (2007).

\bibitem{Ashida}
Y. Ashida, Z. Gong, and M. Ueda,
Non-Hermitian physics,
Adv. Phys. \textbf{69}, 249 (2020).

\bibitem{Palik}
E. D. Palik (ed.),
\textit{Handbook of Optical Constants of Solids}
(Academic Press, Orlando, 1985).

\bibitem{Biehs2021} S.-A. Biehs, R. Messina, P. S. Venkataram, A. W. Rodriguez, J. C. Cuevas and P. Ben-Abdallah,
\emph{Near-field radiative heat transfer in many-body systems}, 
Rev. Mod. Phys., \textbf{93}, 2, 025009 (2021).


\bibitem{pba2011} P. Ben-Abdallah, S.-A. Biehs, and K. Joulain, 
Many-body radiative heat transfer theory, 
Phys. Rev. Lett. \textbf{107}, 114301 (2011).


\end{thebibliography}
\end{document}